# The Memories of the First European Cosmic Ray Symposium: Łódź 1968

The origins of the series of European Cosmic-Ray Symposia are briefly described. The first meeting in the series, on 'Hadronic Interactions and Extensive Air Showers', held in Łódź, Poland in 1968, was attended by the author: some memories are recounted.

**A A Watson** [a,*]

[a] *School of Physics and Astronom, University of Leeds, Leeds, UK*

  *E-mail:* a.a.watson@leeds.ac.uk

[*]Speaker

1.      Introduction

This paper is based on a talk given at the 27th European Cosmic Ray Symposium held in Nijmegen in July 2022.  The first meeting of this series resulted from the efforts of Arnold Wolfendale (University of Durham, UK) and Wlodzimierz (Alexander) Zawadski (Head of the Cosmic-Ray Group in Łódź, Poland), who had met and bonded during the International Cosmic Ray Conference in Jaipur in 1963.  At the height of the Cold War (1947 – 1991), when exchanges were often extremely difficult, they decided to promote links between scientists from Eastern and Western Europe through a series of symposia[1].  In the early years, two biennial meetings were held in different locations at more or less the same time.  One was devoted to High-Energy Interactions and Extensive Air-Showers, while the other was focused on low-energy cosmic rays, and, in particular, on studies of time variations and on work with neutron monitors.  In 1976, the two streams came together for a meeting held in Leeds, UK, to mark the retirement of J G Wilson, a distinguished cosmic-ray physicist who initiated the extensive air-shower project at Haverah Park, and who had worked closely with P M S Blackett.  This format has continued.  Early symposia were held mainly in Western European countries, with the number of Eastern Europeans able to attend gradually increasing.  Around 20 scientists from the USSR came to the Leeds in 1976, and tensions had eased sufficiently for one of them to spend two weeks in the UK after the meeting.

1.      Memories of the meeting

I was fortunate to attend the first event, on 'High Energy Interactions and Extensive Air Showers', held in Łódź, Poland, in April 1968.  Although already 29, this was my first conference outside of the UK.  Nowadays, if you haven't sent your research students on at least a couple of trips abroad during their time with you, you have let them down!  Back then, however, air travel was very expensive and travel generally much more difficult than now, with the complications of visas, travellers cheques etc…..  The next meeting that I attended abroad was four years later, the 3rd European Cosmic Ray Symposium, held in Paris in 1972.  Indeed, I was quite fortunate to be one of the Leeds delegates: my more senior colleagues were unenthusiastic travellers, and, in particular, rather averse to visiting Eastern Europe.

Unfortunately, I now have no notes of the talks given in Łódź, so what follows is focussed on a small area of activity close to my own interests, with my memory prompted by photographs and by the subsequent contacts with scientists from Eastern Europe that were very fruitful, and lasted for many years.  Sadly, none of the friends whom I made at this meeting are still alive.

Comparison of Figure 1 with a modern conference photograph highlights several differences.  Only one woman is present, while all of the men are wearing tie,s and many are in suits!  But, just like today, several people missed the picture.  Only two Russians are visible but at least 5 were at the meeting.  About thirty people attended, approximately half of whom were from Eastern Bloc

---

[1] One of my Leeds colleagues recalls that NATO gave support to these Symposia.  I have been unable to verify this, but certainly NATO support was available for an Advanced Study Institute on the 'Origin of Cosmic Rays' organised by Arnold Wolfendale at the University of Durham in 1974.

countries.

Many years later, I asked Maria Giller, who spent all of her career associated with the Łódź cosmic ray group, and who eventually led it, why she had not been in the photograph. She told me that she was a research student at the time, and so her presence in the photograph had not been permitted. I also found out, again much later, that Maria had been required to change the spelling of her surname to 'Giler' following the Soviet occupation of Poland after WWII. Insights such as this into life in a Soviet-controlled country enriched the experiences derived from the visit.

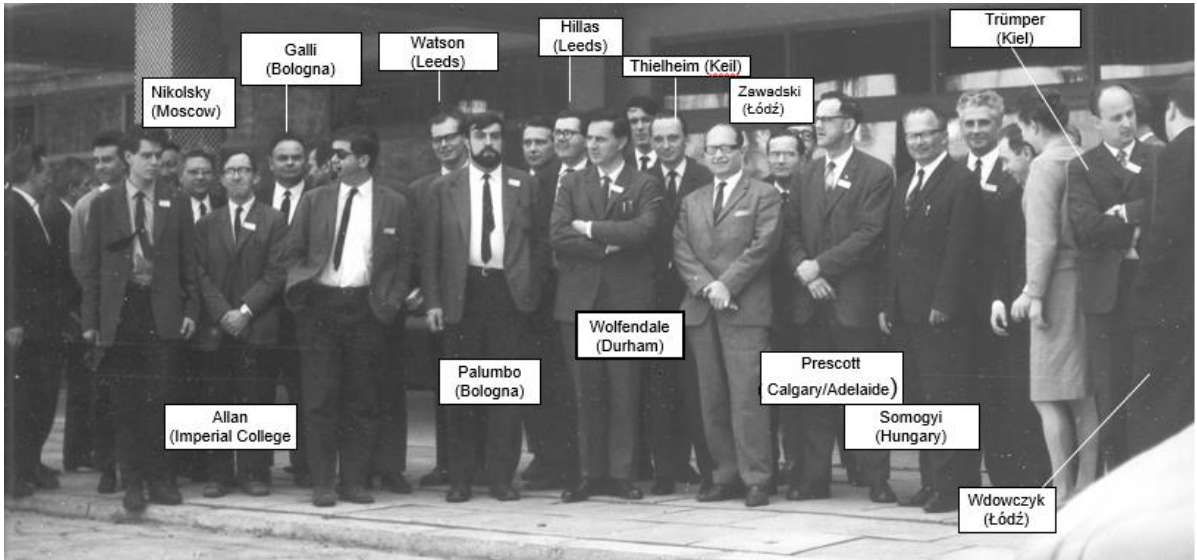

Figure 1: Some of those who attended the first European Cosmic Ray Symposium.

The meeting remains in my memory as one of the most productive that I have ever attended. All talks were of essentially unlimited length and were given in English or Russian, with translation (not simultaneous) by one or other of the participants. This contrasted with what I'd seen at the International Cosmic Ray Conference held in London in 1965 where senior Russians, such as G T Zatsepin, raced from session to session, giving talks on behalf of colleagues who were present, but did not speak English. Sometimes Zatsepin did not even agree with the arguments advanced in the papers that he was presenting! Also the Russian 'minders', so evident in London, seemed to be absent. There was plenty of time for discussions so that each of the five days of the meeting was very long.

2. Some of the topics discussed

Nowadays, water-Cherenkov detectors are used in many projects in Astroparticle Physics. This was not the case in 1968. Although Chudakov, in the Lebedev Institute, had proposed a relatively large water-Cherenkov detector in the early 1950s, true to form, he never wrote up a description of his 85 tonne detector. It was in the form of a truncated cone, viewed by 16 photomultipliers with 6 inch photocathodes, and was used in a study of muon bundles in 1963 [1]. A water-Cherenkov detector had been developed in the UK at around the same time by Porter et al. [2], working at the large (~0.6 $km^2$) array of Geiger counters at Harwell. Allan et al. [3] had subsequently used similar detectors in a study of the knee in the shower number spectrum, first reported by Kulikov and Khristiansen [4] who worked exclusively with Geiger counters.

However, at the time of the Łódź meeting, Geiger counters were almost the only detectors available to scientists in Eastern Europe. Consequently, the use at Haverah Park of only water-Cherenkov detectors (over 200 tanks each containing ~3 tonnes of water) aroused much suspicion and criticism, and so one of my tasks was to try to explain and justify what we were doing. At the time, Geiger counters were used to give the number of charged particles in a shower, and it was expected that results from arrays using scintillation counters (for example those developed by the MIT group) would be converted to numbers of charged particles to follow tradition and aid comparisons. To a limited extent, this was possible with scintillators as they are relatively insensitive to photons, but it was completely impractical with water-Cherenkov detectors. There are about ten times as many photons in showers as there are electrons, and these were totally absorbed in the 3.4 radiation lengths of the Haverah Park detectors. Photographs of the author trying to explain these points are in figure 2. Note the use of chalk – overhead projectors and acetate sheets were still some years off, with PowerPoint not even a dream.

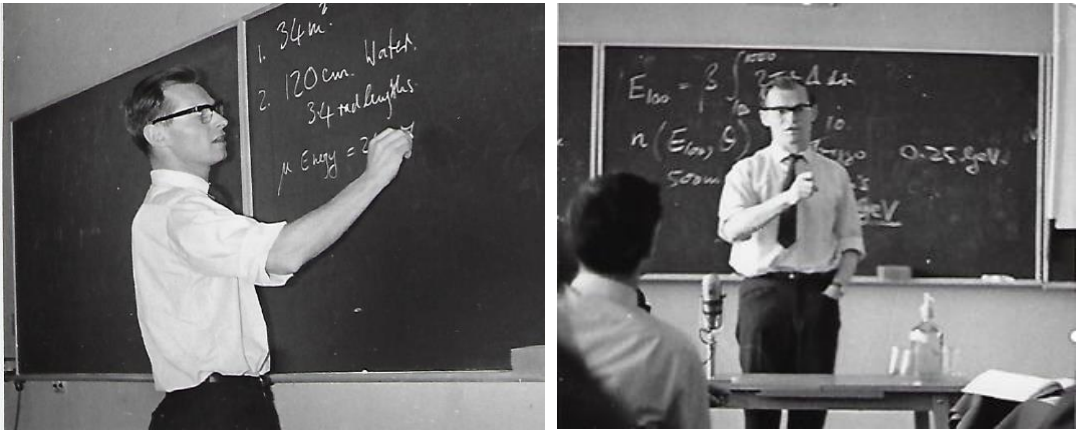

Figure 2: The author giving a talk about the water-Cherenkov detectors used at Haverah Park.

In 1966, Greisen, and Zatsepin and Kuzmin had pointed out that an implication of the existence of the cosmic microwave background radiation was that there would be a sharp fall in the flux of cosmic rays at energies above ~50 EeV. A few months before the Łódź meeting, the Haverah Park group had recorded an event for which the energy estimate was above this limit. I described this event, and details were published a few months later [5]. Re-reading this paper, with the benefit of over 50 years further experience in the field is rather embarrassing, as we made a number of assumptions that are now impossible to defend. But, it was an exciting time and the event was seen as important as this was the first such energetic particle detected since that reported by Linsley [6] five years earlier. The paper appeared in print only 24 days after submission[2].

Another topic that got much attention during the meeting was the possibility of using radio emission from air showers to build large arrays of detectors economically, and to study features of shower development. Harold Allan, although not involved in the first detections of radio emission, had quickly become one of the experts in the field and is seen describing the essence of radio production in figure 3. Harold was a superb scientist and a brilliant communicator.

---

[2] Between 1939 and 1973, papers submitted to Nature were not routinely subjected to Peer Review: see M Baldwin, Notes and Records of the Royal Society 69 37 2015.

It is perhaps ironic that Giorgio Palumbo (figure 1), and who became a good friend, was to go on to show that it was, at that time, impractical to use radio studies to measure the depth of shower maximum. His paper [7] helped to shut down this field of work for more than 30 years.

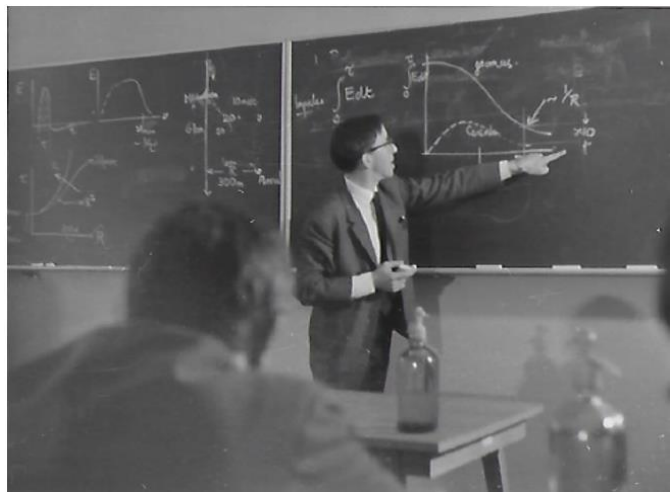

Figure 3: Harold Allan lecturing on radio emission from Extensive Air Showers

3. Networking opportunities

There were many opportunities for networking during the meeting. In figure 3, Harold Allan can be seen, flanked by four of the Russians who were present, S I Nikolsky, A D Erlykin, N N Efimov and G B Khristiansen.

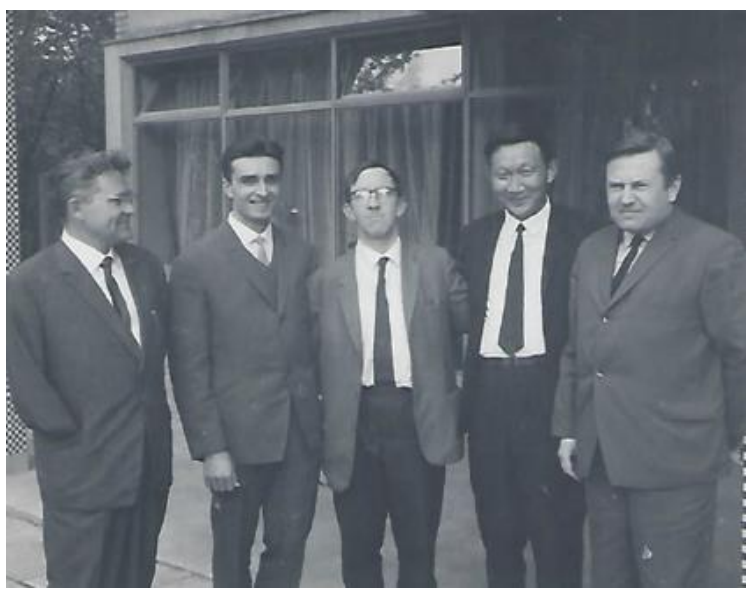

Figure 4: Harold Allan with S I Nikolsky, A D Erlykin, NN Efimov and G B Khristiansen

Partly as a result of meeting Nikolsky, I was invited to attend a conference at the Lebedev Institute in Moscow a few months later. This event was held to mark his appointment as one of the Vice-Chairmen of the Institute, and while I was there, Anatoly Erlykin was one of a number of Muscovites who made me welcome. George Khristiansen became a good friend – a first-rate physicist and a very good table-tennis player. From Nick Efimov, I learned of the large shower array in Yakutsk which had just started taking data, and which I visited for two weeks in 1984, when he was a most gracious host. Harold Allan subsequently visited Moscow to discuss radio issues with Khristiansen on several occasions, while I made five visits to the USSR, and a further two visits after Glasnost. These visits would been unlikely to have taken place had I not attended the Łódź meeting.

Even before this meeting, Arnold Wolfendale had begun what was to become a long standing and very productive collaboration with Jerzy ('George') Wdowczyk and other members of the Łódź group. They are seen in animated discussion in figure 4.

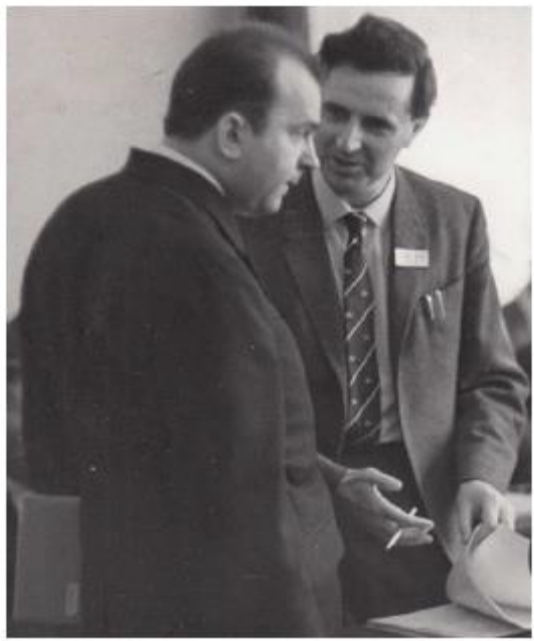

Figure 4: George Wdowczyk and Arnold Wolfendale in discussion during the Łódź meeting

4.      Aspects of Polish Culture

Our Polish hosts were keen to look after our cultural health as well as offering an exciting physics program. One afternoon, we were taken to an old church, some distance from Łódź, where we were treated to a Chopin piano recital by candlelight. The trip also gave us a chance to see the limited mechanisation on the nearby farms – horses rather than tractors – and other aspects of the infrastructure. One evening we were taken to a performance of Borodin's Prince Igor, the dance of the Polovtsian Maidens being particularly erotic.

After the meeting ended, we visited Warsaw for sightseeing before our flights home. We stopped off at Vilanova Castle, but most memorable was the visit to the Old Town, meticulously rebuilt following its total destruction during World War II. For this, we were divided into three groups: Russians and other Eastern Europeans, Germans, and Others. No doubt, the guides told rather different stories about the destruction of this area, now so popular with tourists.

6.     Other Memories

In the 1960s, getting flights to Eastern Europe, and acquiring the necessary visas, was not straightforward. The Daily Worker, a British newspaper funded by the Communist Party, was the usual starting point, but sending off one's money and passport to an unknown travel agent, and then waiting for the tickets and the visa to arrive was a nail-biting experience. For some strange reason, we had to change planes in East Berlin, and I recall a long wait in a windowless room in East Berlin airport, the walls decorated only with a photograph of Walter Ulbricht, Chairman of the State Council of the German Democratic Republic. I was reminded of this when I reached the South Pole some 20 years later, where a picture of President Ronald Regan greeted one at the entrance to the main building.

In the hotel, several of us got late night-calls from the Reception Desk asking if we would like a young lady to keep us company. When that failed, one was offered a young man instead.

I had a number of Polish friends in Leeds and visited the Polish Club there fairly often. One had asked me to take some gifts to his friends in Łódź, which I gladly did. They welcomed me warmly to their home, fed me good food and introduced me to Zubrowka. I was very fortunate to have this opportunity. While in Łódź and Warsaw, there were chances to wander around the streets, unaccompanied, so different from what I was to find in later in Moscow. During these wanderings, if I was with Giorgio Palumbo, we were frequently asked if we would like to change money. Perhaps Giorgio's beard sent out some sort of signal that was lacking for, when I was alone, I was never approached.

In hindsight, this meeting was very important to me, for the science, the contacts and the friends whom I made there. Nothing comparable could have been experienced over Zoom.

Acknowledgments

I would like to thank Jörg Horandel for inviting me to give the talk on which this paper is based, and for the financial help that made my attendance possible. It was a first-class meeting with some excellent discussions: the oysters served at dinner one evening are particularly memorable!